\documentclass[aps,pra,twocolumn,showpacs,amsmath,amssymb,preprintnumbers,superscriptaddress,10pt]{revtex4-2}
\usepackage{graphicx}
\usepackage{subfig}
\usepackage{siunitx}
\graphicspath{{figures/}}
\usepackage{ragged2e}
\usepackage{hyphenat}
\usepackage{braket}
\usepackage[bookmarks=false]{hyperref}
\usepackage{color}
\usepackage{verbatim}
\usepackage[capitalise]{cleveref}

\usepackage{float} 

\usepackage[section]{placeins}
\begin{document}
\title{High-efficiency Weak-trace-free Counterfactual Communication via Quantum Zeno Effect}

\author{Tianyi Xing}
\thanks{These authors contributed equally}
\affiliation{College of Computer Science and Technology, National University of Defense Technology, Changsha 410073, China}

\author{Anqi Huang}
\thanks{These authors contributed equally}
\email{angelhuang.hn@gmail.com}
\affiliation{\mbox{College of Electronic Science and Technology,} \mbox{National University of Defense Technology, Changsha 410073, China}}
\affiliation{Center for Cryptologic Research, National University of Defense Technology, Changsha 410073, China}
\affiliation{College of Computer Science and Technology, National University of Defense Technology, Changsha 410073, China}

\author{Yizhi Wang}
\affiliation{College of Computer Science and Technology, National University of Defense Technology, Changsha 410073, China}

\author{Chao Wu}
\affiliation{College of Computer Science and Technology, National University of Defense Technology, Changsha 410073, China}

\author{Yaxuan Wang}
\affiliation{College of Computer Science and Technology, National University of Defense Technology, Changsha 410073, China}

\author{Pingyu Zhu}
\affiliation{College of Computer Science and Technology, National University of Defense Technology, Changsha 410073, China}

\author{Jiangfang Ding}
\affiliation{College of Computer Science and Technology, National University of Defense Technology, Changsha 410073, China}

\author{Dongyang Wang}
\affiliation{College of Computer Science and Technology, National University of Defense Technology, Changsha 410073, China}

\author{Yingwen Liu}
\affiliation{College of Computer Science and Technology, National University of Defense Technology, Changsha 410073, China}

\author{Xiaogang Qiang}
\affiliation{National Innovation Institute of Defense Technology, AMS, 100071 Beijing, China}

\author{Sheng Ma}
\affiliation{College of Computer Science and Technology, National University of Defense Technology, Changsha 410073, China}

\author{Ping Xu}
\email{pingxu529@nju.edu.cn}
\affiliation{College of Computer Science and Technology, National University of Defense Technology, Changsha 410073, China}

\author{Junjie Wu}
\affiliation{College of Computer Science and Technology, National University of Defense Technology, Changsha 410073, China}

\date{\today}

\begin{abstract}

The quantum Zeno effect, which inhibits quantum state evolution via repeated weak measurements, significantly enhances the efficiency of interaction-free measurement (IFM). This fundamental mechanism facilitates high-efficiency counterfactual quantum communication, enabling information delivery without particle transmission through the channel. However, the transmission time of the counterfactual communication requires minutes for bit and suffers the bit error when transmitting an image. Applying the quantum Zeno effect, we experimentally demonstrate high-efficiency weak-trace-free counterfactual communication on a quantum photonic chip, achieving a transmission probability of 74.2 ± 1.6\% for bit 0 and 85.1 ± 1.3\% for bit 1. Furthermore, we successfully transmit our group's logo -- Quanta -- through counterfactual communication, and reduce the time cost from minutes to seconds for bit, with zero bit errors after information processing. Our study provides a promising approach for secure and efficient communication using integrated silicon quantum photonics.

\end{abstract}
\maketitle

{\bfseries keywords}: counterfactual communication, quantum Zeno effect,  quantum photonic integration

\section{Introduction}
\label{sec:intro}

Quantum Zeno effect, as a fundamental concept of quantum mechanism, inhibits the evolution of quantum states through frequent but weak measurements. It can be likened to an arrow appearing to be in flight at any given moment without actually moving~\cite{misra1977zeno}. This inhibition occurs because the collapse of the wave function back to its initial state is induced by short-term measurements. This paradoxical feature provides a means to freeze the initial state, allowing a quantum system coupled with a measurement device to maintain high fidelity quantum states through a series of weak observations~\cite{itano1990quantum}.

The quantum Zeno effect can be utilized to significantly enhance the efficiency of an interaction-free measurement (IFM) that detects an obstructing object placed in an interferometer without any direct interaction with an interrogating particle, owing to the wave-particle duality inherent in a quantum system~\cite{Dicke1981Interaction,elitzur1993quantum}. By employing repeated weak measurements as a manifestation of the quantum Zeno effect, it ensures that the initial quantum state of the interrogating particle is maintained with a high probability, thereby greatly improving the detection efficiency of the obstructing object from 33\% to approach near-perfect levels close to 100\%~\cite{Paul1995Interaction}.

The high-efficiency IFM is employed in various applications, such as quantum interrogation~\cite{1999High}, quantum tomography~\cite{potting2000quantum,facchi2002Quantum,elouard2020interaction}, quantum computation~\cite{paz2012zeno,hosten2006counterfactual,peise2015interaction,qiang2021implementing}, and quantum cryptography~\cite{shukla2014orthogonal,li2016quantum}. Importantly, this physical phenomenon is highly suitable for secure communication known as counterfactual communication, wherein information is conveyed without the transmission of any particles through the channel. As shown in Fig.~\ref{Fig:Schematic}(a), in classical communication, the communicating parties (Alice and Bob) must transmit information via physical particles passing through the channel, which may be eavesdropped by Eve. In counterfactual communication, however, Alice emits a quantum state to herself, and Bob conveys information by altering the evolution outcome of this state via $\hat{P}$. Throughout the entire process, no particles pass through the transmission channel. Thus the eavesdropper Eve cannot intercept any particles from the channel, and consequently cannot obtain the information. Furthermore, by utilizing the quantum Zeno effect, the success rate of counterfactual communication can be increased to 100\% in theory~\cite{Hatim2013}.

A practical protocol for achieving direct counterfactual quantum communication was proposed in Ref.~\cite{Hatim2013,salih2022laws}. Subsequently, a free-space optical experimental demonstration of this protocol was conducted to transmit a Chinese knot, thereby demonstrating its feasibility~\cite{cao2017direct}. However, it was discovered that when delivering bit 0, the protocol failed to maintain complete counterfactual information delivery and left behind a trace of the photon in the transmission channel~\cite{vaidman2014comment}. Fortunately, an upgraded weak-trace-free protocol was developed in 2019 to ensure full counterfactuality by utilizing the two-state vector formalism (TSVF) as evidence that no photons leaked into the transmission channel~\cite{aharonov2008two,aharonov2019modification}. Recently, a basic implementation of this modified protocol using bulk optics has been demonstrated~\cite{pan2023counterfactual} and successfully transmit a QR code. However, due to the absence of the quantum Zeno effect, this implementation may face security threats, has limited success probabilities, and incurs transmission time overhead~\cite{pan2023counterfactual}.

\par

In this study, we leverage the quantum Zeno effect to experimentally demonstrate weak-trace-free counterfactual communication on a quantum photonic chip and illustrate its counterfactual property in information delivery by transmitting the logo of our Quanta group. Our demonstration takes advantage of the compact and stable on-chip optical platform, achieving the success probabilities of $74.2 \pm 1.6\%$ for bit 0 and $85.1 \pm 1.3\%$ for bit 1 in counterfactual transmission. Furthermore, only an average of $0.17\%$ photons are detected in the transmission channel due to practical interference imperfections, highlighting the highly counterfactual property of bits' transmission. Notably, we cut the one-bit transmission time from minutes to seconds via this high-efficiency counterfactual communication. In the scenario of 1 second per bit, it only takes 42 minutes in total to transmit our ``Quanta" group's 50×50 - pixel grayscale logo. Moreover, by setting the one-bit transmission time 5 seconds, we achieved error-free and low-noise transmission.

Our study represents the first experimental demonstration of the weak-trace-free counterfactual communication protocol based on the quantum Zeno effect and TSVF. Utilizing a programmable and stable quantum photonic chip, we physically realize the necessary inner and outer cycle chains of interferometers to achieve the quantum Zeno effect. This enables us to enhance the success probability of information transmission and reduce one-bit transmission time from minutes to seconds. Additionally, we implement the extra but essential inner cycles of interferometers as proposed by the modified protocol, ensuring no weak trace left in the transmission channel. This property is verified in our experiment, indicating that counterfactuality is maintained for both bit 0 and bit 1, in contrast previous work leaving a loophole of weak trace~\cite{cao2017direct,alonso2019trace}. This research indicates a potential approach to achieve secure communication with high efficiency through integrated silicon quantum photonics.\par

\section{Result}
\label{sec:Result}

\subsection{Counterfactual communication protocol}
\begin{figure*}[t]
    \centering
\includegraphics[width=0.9\textwidth]{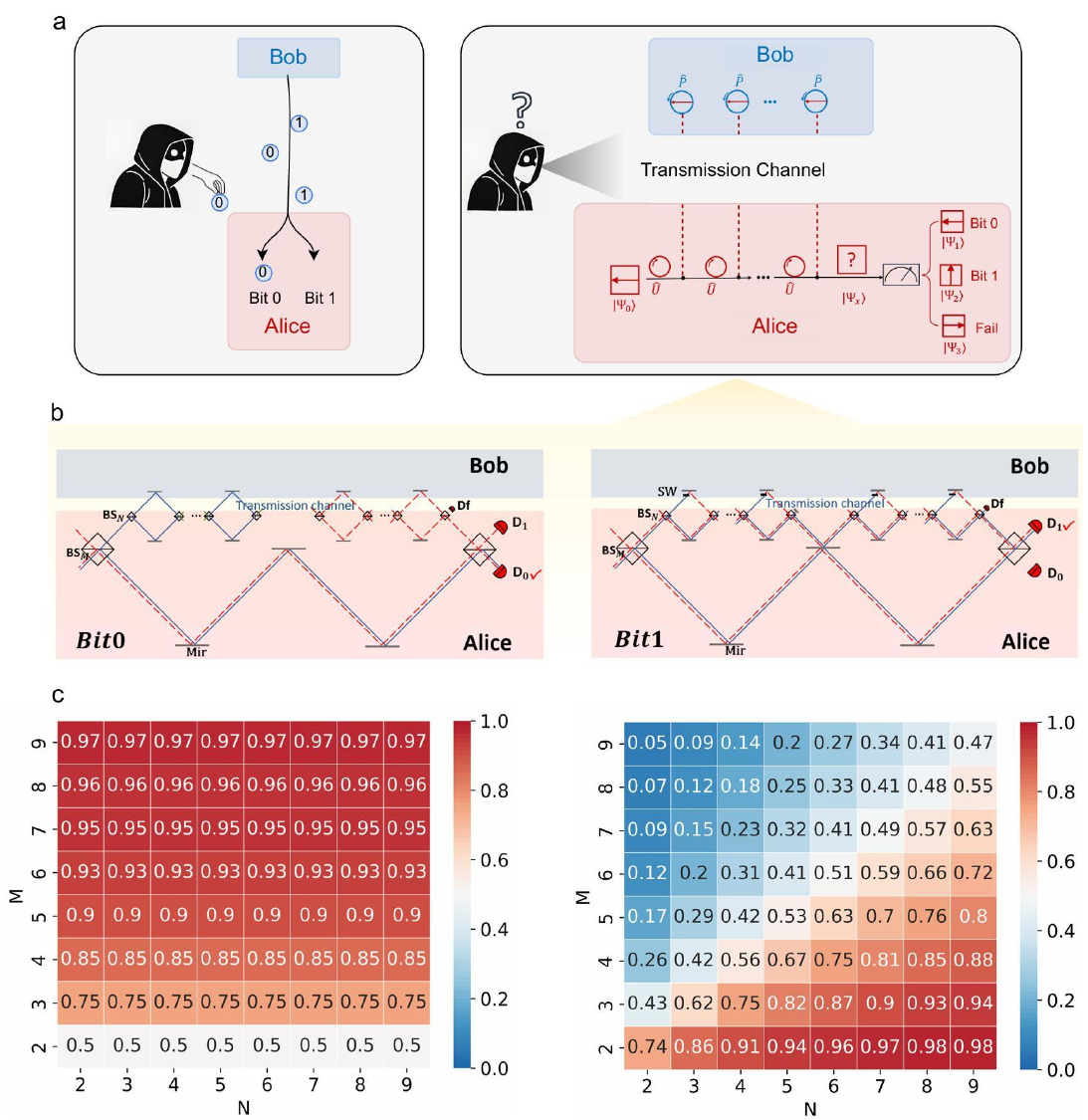}
   \caption{\justifying{{\bfseries The schematic and structure of the upgraded counterfactual communication protocol.} (a) During the classical communication process, Eve can obtain the information from Alice to Bob by intercepting the information-carried particles. However, since Alice and Bob do not transmit any particles during the counterfactual communication, Eve can not acquire any information. (b) The updated protocol for the counterfactual communication. (c) The success probabilities of transmitting bit 0 and bit 1 are simulated with various $M$ and $N$. Balancing the trade-off between feasibility of implementation and success probabilities of transmission, we choose $M = 3$ and $N = 6$ in our experimental scheme.}}
   \label{Fig:Schematic}
\end{figure*}

In the classical communication, Alice encodes the information in the particles and transmits them to Bob. Eve can intercept these particles and obtain the information from Alice, which leads to security risk. In
contrast, counterfactual communication offers an approach to information exchange without particles transmission, as shown in Fig.~\ref{Fig:Schematic}(a). In the counterfactual communication, Alice applies $\hat{U}$ that drive the the initial quantum state $\ket{\Psi_0}$ into $\ket{\Psi_2}$ or $\ket{\Psi_3}$. Bob controls the adjustable projectors (${\hat{P}}$) that keep the state staying at $\ket{\Psi_1}$ with ${\hat{P}_1}=\ket{\Psi_1}\bra{\Psi_1}$, or evolve to $\ket{\Psi_2}$ with ${\hat{P}_2}=\ket{\Psi_1}\bra{\Psi_1}+\ket{\Psi_2}\bra{\Psi_2}$. If the state evolves into $\ket{\Psi_3}$, the counterfactual transmission fails. Finally the state is measured by Alice as $\ket{\Psi_x}$, with $\ket{\Psi_1}$ for bit 0, $\ket{\Psi_2}$ for bit 1 and $\ket{\Psi_3}$ for failure. The repeated structure with adjustable projects can realize quantum Zeno effect and transmit the information from Bob to Alice.

The quantum Zeno effect that inhibits transitions between quantum state can be clearly explained in the context of a two-level system. Generally, the wave function of a two-level system interacting with an external field can be written as~\cite{cook1988quantum, home1993unified, home1992negative}
\begin{equation}
\ket{\psi(t)} = cos(\frac{1}{2}\Omega t) \ket{0} + e^{i\phi}sin(\frac{1}{2}\Omega t) \ket{1},
\end{equation}
where $\phi$ is a phase factor, and $\Omega$ is the Rabi frequency of the applied field that excites the state transition. Thus, after a period of T, the state undergoes a complete transition to the excited state, $\ket{\psi(T)} = e^{i\phi}\ket{1}$, in which $\Omega T = \pi$. Considering the first intermediate measurements at time $T/N$, the probability of detecting at the ground state becomes
\begin{align}
P_{0}(T/N) & = cos^2(\Omega T/2N) = cos^2(\pi/2N) 
\\&\approx 1-\pi^2/4N^2,
\end{align}
It is obvious that after N times of measurement, 
\begin{equation}
P_{0}^N(T/N) = cos^{2N}(\pi/2N) \approx (1-\pi^2/4N^2)^N \rightarrow 1, 
\end{equation}
if the measurement required by the quantum Zeno effect is repeatedly applied in every short period $T/N$ as $N \rightarrow \infty$.

Quantum Zeno effect can be realized by the optical scheme as well~\cite{agarwal1994all}. The operation of an optical beam splitter (BS) with reflection coefficient of $cos\theta$, playing the same roles as $cos(\frac{1}{2}\Omega t)$. Thus, the probability of detecting the reflected single-photon state after passing through a BS is $P = cos^2\theta$. If an obstructing object is introduced into an interferometer used for IFM, it disrupts the coherence of the quantum state and projects the system back to its ground state. For tandom interferometers used in the IFM, the detection probability becomes $P = cos^{2N}\theta$. Set the reflection coefficient to be $\theta = \pi/2N$, 
\begin{equation}
P = cos^{2N}\theta = cos^{2N}(\pi/2N) \rightarrow 1.
\end{equation}
The quantum Zeno effect is achieved through the repeated implementation of IFM, thereby significantly improving its efficiency~\cite{Paul1995Interaction}.

Therefore, the specific optical scheme and the operation of the upgraded protocol are as follows, shown in Fig.~\ref{Fig:Schematic}(b). The outer cycles of interferometers consist of $M$ beam splitters~($BS_M$s) with the reflectivity of $cos^2(\pi/2M)$ each. Serving as one arm of the outer interferometer, the inner cycles contain two sets of tandem interferometers connected by a two-sided mirror. Each set of inner interferometers comprises $N$ beam splitters~($BS_N$s) with a reflectivity of $cos^2(\pi/2N)$ for each $BS_N$. In accordance with the quantum Zeno effect, the photon state remains stable within the nested cycles of interferometers. The structural unit of the trace-free protocol is illustrated in Fig.~\ref{Fig:Schematic}(b) featuring one outer cycle and corresponding inner cycles. Additionally, as shown in Fig.~\ref{Fig:Schematic}(b), two single-photon detectors~(SPDs), denoted as $D_0$ and $D_1$, are positioned at the end of Alice's outer interferometer cycles to detect bit 0 or 1 without photons traversing across the channel, while detector $D_f$ is placed at the end of the inner cycles to observe whether photons enter into the transmission channel.

The information transmitted is determined by Bob's operation of switches~(SWs), which act as obstructing objects. If Bob closes the switches to obstruct the evolution, choosing ${\hat{P}}={\hat{P}_1}$ in Fig.~\ref{Fig:Schematic}(a), the quantum state of the photon is kept as initial and will be detected by $D_0$ for bit 0. Conversely, if Bob opens the switches to allow the photon to pass, choosing ${\hat{P}}={\hat{P}_2}$ in Fig.~\ref{Fig:Schematic}(a), its quantum state evolves through interference and will be detected by $D_1$ for bit 1. As the numbers of $M$ and $N$ increase, the probability of successful transmission approaches 100\% in theory due to the quantum Zeno effect, as shown in Fig.~\ref{Fig:Schematic}(c).

Furthermore, the repeated inner tandem interferometers prevent the overlap of forward evolving waves in yellow and backward evolving waves in red, as described by TSVF \cite{aharonov1964time,aharonov2008two}. Consequently, the overlapping part of the yellow and red lines represents the trace of the photon during transmission in reality. It is important to note that no trace of the photon remains in the transmission channel to ensure full counterfactuality, a concept which will be experimentally validated through observation by $D_f$. In summary, this upgraded counterfactual communication enables information to be transmitted from Bob to Alice indicated by clicks on SPDs ($D_0$ or $D_1$), while ensuring that photons never physically travel between Alice and Bob.

\subsection{Experimental implementation}
Although the probability of detecting the correct bit approaches 100\% under infinite cycles of the outer and inner interferometers, it is impractical to realize extremely large numbers of $M$ and $N$ in an experiment. The core of the experimental design lies in selecting feasible values for $M$ and $N$, which determine the overall structure and success probabilities. In the modified protocol for counterfactual communication, a larger value of $M$ results in a lower error probability for transmitting bit 0, as given by $1-cos^2(\pi/2M)$. Regarding bit 1, the probability of transmission error depends on both values of $M$ and $N$. Given that larger values of $M$ and $N$ lead to a more complex structure with reduced interference visibility and increased channel loss, it is essential to identify appropriate values for $M$ and $N$ to maintain relatively high success probabilities while ensuring stable interference with acceptable channel loss. The success probabilities for transmitting bits 0 and 1 are simulated according to the modified protocol for counterfactual communication, as shown in Fig.~\ref{Fig:Schematic}(c), where $S_0$ represents success probability for bit 0 and $S_1$ represents those for bit 1.

Figure~\ref{Fig:Schematic}c illustrates the rapid increase of $S_1$, the success probability of bit 1, with an increase in $N$ for each $M$, followed by a saturation range. This indicates that there exists an optimal value of $N$ approaching saturation for each $M$, achieving a relatively high success probability with a relatively small number of $N$. However, it is important to note that there are $(M - 1) \times (N-1)$ Mach-Zehnder interferometers~(MZIs) for photons to travel before being detected, leading to geometrically increasing hardware cost and channel loss as $M$ and $N$ increase. As such, there is a limitation on how large M and N can be. As a trade-off between the detection probabilities and experimental complexity, we have chosen $M = 3$ and $N = 6$ in our experimental implementation, which theoretically yields success probabilities of 75.0\% for bit 0 and 87.2\% for bit 1 respectively.

\begin{figure*}[t]
    \centering
    \includegraphics[width=0.9\textwidth]{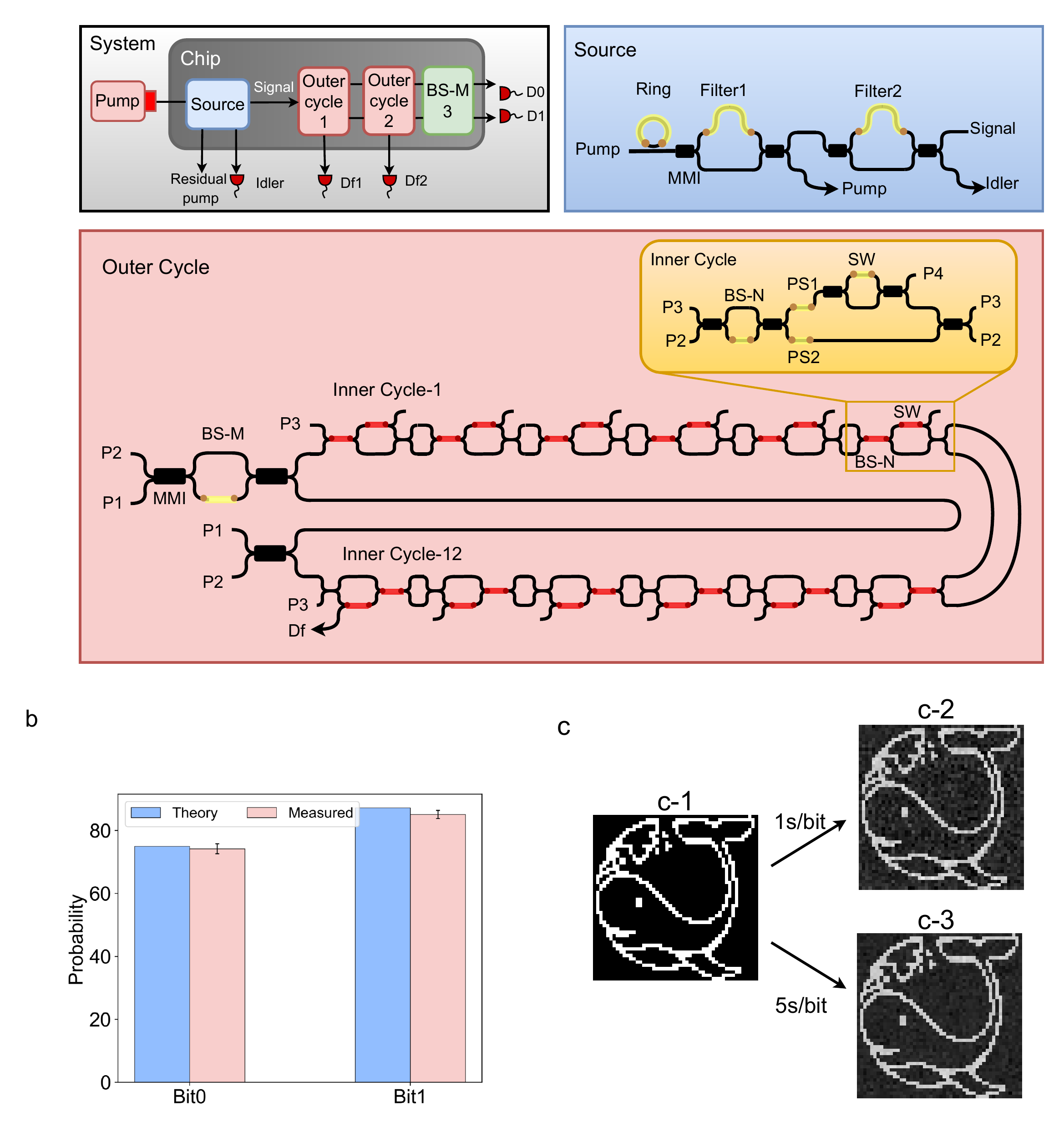}\caption{{\bfseries \bfseries The experimental setup and transmission result.} \justifying{(a) The chip contains photon-pairs source and the counterfactual transmission structure. The structures of outer and inner cycle in chip are shown as above, and the two red components in the outer cycle represent the reconfigurable BS-N and the SW in the inner cycle. (b) Success probabilities for bit 0 and bit 1 are 74.2 ± 1.6\% and 85.1 ± 1.3\%, compared to theoretical limits. (c) The transmission of our group logo, Quanta, in  $50 \times 50$ pixels. We transmitted the logo in different transmission time cost for 1 bit, with $T$ =\SI{1}{\second} for 1 bit in (c-2) and $T$ =\SI{5}{\second} for 1 bit in (c-3). }}\label{fig:Layout}
\end{figure*}

Following the upgraded protocol and the choice of M/N, as shown in Fig.~\ref{fig:Layout}(a), the counterfactual communication implemented on the quantum photonic chip is structured with outer and inner cycles, providing four possible paths for the input photon (P1 to P4). The input photon is generated by a heralded single-photon source on the chip, which includes a ring resonator and two asymmetric MZIs serving as filters. In terms of the counterfactual transmission component, two outer cycles are configured with $M = 3$ in the chip layout, consisting of three beam splitters ($BS_M$s) with a reflectivity of $cos^2(\pi/2M)$. One arm of each outer cycle includes two repeated structures --- five inner cycles (twelve $BS_N$s in total with $N = 6$), ensuring the counterfactuality of transmission. As illustrated in Fig.~\ref{fig:Layout}(a), each inner cycle primarily consists of two adjustable MZIs: one functioning as a $BS_N$ in paths 2 and 3, while the other acts as a SW  in path 3 and 4 controlled by Bob to either allow or block photons from entering path 3. Additionally, two adjustable phase shifters (PSs) are utilized in each cycle to optimize interference: PS1 compensates for phase offset caused by $BS_N$ in path 3, while PS2 serves as a spare element in path 2 to balance phase. Following the transmission scheme, Alice connects SPDs $D_0$ and $D_1$ at end of entire structure to detect information corresponding to bit 0 and bit 1 respectively. Meanwhile, an additional pair of SPDs, $D_{f1}$ and $D_{f2}$, are employed at the final inner cycle of each outer cycle to validate the counterfactual property ---- specifically, whether the photons actually traverse through the channel from Alice to Bob in practical application. The complete experimental setup is shown and detailed in~\cref{sec:Method}.

In addition to the systematic scheme and on-chip structure mentioned above, another critical factor for achieving the protocol is the precise control of phases applied by all the PSs on the chip. To accomplish this, each PS is pre-calibrated to establish the relationship between the phase angle and the applied voltage, based on the thermal-optical effect. After these calibrations, we characterize the interference visibility of 27 MZIs in the cascade. The results show visibilities ranging from 99.42\% to 99.99\%, with a mean of 99.88\% and a standard deviation of 0.0013, demonstrating excellent and consistent performance across all interferometers. Furthermore, due to variations in the number of multi-mode interferometers (MMIs) in the two arms of tandem interferometers, phase compensation and loss offset are implemented for these arms (further details in Sec.~IV). This optimization of phase implementation ensures accuracy in control and precision in the success probability of information transmission for counterfactual communication. After global loss and phase compensation, the overall interference visibility of the entire structure is measured at a representative operating point to be 99.67\%, confirming that the system achieves near-unity visibility under optimized conditions.

In the case of transmitting bit 0, Bob opens the SW, which is realized by an MZI placed between path 3 and path 4 of each inner interferometer on the chip. The photon is then allowed to pass the SW and is detected by $D_0$. In the case of transmitting bit 1, Bob closes the SW to block the photon that is finally detected by $D_1$. Theoretically, in the scheme of modified counterfactual communication with $M = 3$ and $N = 6$, the probabilities of information transmission from Bob to Alice are $75.0\%$ for logic 0 and $87.2\%$ for logic 1. In our implementation, these identification probabilities are $74.2 \pm 1.6\%$ and $85.1 \pm 1.3\%$ in Fig.~\ref{fig:Layout}(b), respectively. Meanwhile, the conditional detection rate of $D_f$s and $D_0$ / $D_1$ is $0.17\%$ in average ($0.09\%$ for $D_0$ and $0.25\%$ for $D_1$), one order of magnitude lower than the previous results with weak trace leaked to the channel~\cite{cao2017direct}.

We use this setup to transmit the logo of our group named QUANTA, as shown in Fig.~\ref{fig:Layout}(c-1), in a $50 \times 50$ pixel grayscale chart, and demonstrate the direct counterfactual communication result as shown in Fig.~\ref{fig:Layout}(c-2) and Fig.~\ref{fig:Layout}(c-3). Bob sent bits by controlling the SWs to block or pass the transmitted photons with a total channel loss of \SI{11}{\dB} for bit 0 and \SI{17}{\dB} for bit 1. To guarantee the accurate detection probability in Alice, each bit in Fig.~\ref{fig:Layout}(c-2) is repeatedly transmitted for \SI{1}{\second}. Then bit value of each pixel is sent sequentially and the transmission of the whole diagram takes 42 minutes, which is much faster than the previous work with 1000 seconds for one bit~\cite{pan2023counterfactual}. When Alice obtains detection for each bit, she calculates the correct detection probability and converts the pixel value into grayscale in 5 hours. 

During pixel value converting, the successful probabilities are linearly mapped, as 100\% bit 0 to grayscale 0, and 100 \% bit 1 to grayscale 255. In other words, the shade of gray indicates the transmission success rate of each bit. Figure~\ref{fig:Layout}(c-2) indicates that our counterfactual communication at a rate of 1 bit/s results in some noise in the transmission. It is notable that as the one-bit transmission time ($T$) increases from \SI{1}{\second} to \SI{5}{\second}, the transmission outcome can be more stable to present a less-noisy chart as shown in Fig.~\ref{fig:Layout}(c-3), without bit flip and thus no error. The result of the transmission shown in Fig.~\ref{fig:Layout}(c) indicates that the information in this chart is successfully transmitted from Bob to Alice, under the modified counterfactual communication protocol, without the photon transmitted through the transmission channel.

\section{Discussion}
In theory, the original protocol has the problem that some photons will enter the $D_f$ when $D_1$ or $D_0$ clicks for finite $M$ and $N$. The modified protocol has already solved the loophole of counterfactual by appending two series of inner cycles in the outer cycle. Whereas, we still set a detector $D_f$ to verify the counterfactuality of our implementation, which meanwhile can measure the imperfect interference in the inner cycles. In the case of logic 0, the photons are allowed to pass the inner cycle and are dropped in the upper part of the inner cycle, which means they will not be detected by $D_f$ theoretically. In the case of logic 1, the photons are blocked by Bob, so they can not enter the transmission channel. 

Specifically, in the experiment, the SNSPDs, $D_{f1}$ and $D_{f2}$, are placed at the end of each inner cycle as Fig.~\ref{fig:Layout}(a) shown to check whether the photons entering the transmission channel in practice. We consider the click of $D_{f1}$/$D_{f2}$ while $D_0$ ($D_1$) clicks simultaneously as a violation of counterfactual property. The coincident counts of $D_{f1}$ / $D_{f2}$ and $D_0$ ($D_1$) are summed as the detection rate of $D_f$ for bit 0 (bit 1). This coincidence counts utilizes raw detection data without subtracting dark counts. This ensures the reported $D_f$ detection rate remains a conservative upper bound, explicitly accounting for potential background noise within the results. The experimental results illustrate that the detection rate of $D_f$ is $0.17\%$ in average ($0.09\%$ for bit 0 and $0.25\%$ for bit 1 respectively) while the transmission rate achieved in our experiment is $74.2\%$ for logic 0 and $85.1\%$ for logic 1.

In our implementation of counterfactual communication, the high rate of $D_0$/$D_1$ while low detection of $D_f$s is achieved by the long cycle interferometers, which matches well with the theoretical protocol. The experimental success probabilities—$74.2\%$ for bit 0 and $85.1\%$ for bit 1—are only slightly lower than the theoretical targets of $75.0\%$ and $87.2\%$. This minor discrepancy is attributed to imperfect loss compensation. Although we compensate for the total measured path loss to be almost balanced across repeated MZI structures (see Sec.~\ref{compensation} for details), fabrication variances mean that individual MMIs exhibit unique loss profiles. This prevented perfect rebalancing, resulting in the small deviations between the experimental results and their theoretical counterparts.

According to the transmission result of $D_0$/$D_1$ and $D_f$s, both for bit 0 and bit 1, no weak trace is left in the transmission channel while the information from Bob is carried by the particles in Alice. The channel capacity theory indicates that only one bit of information can be transmitted as the maximal data rate when one photon is transmitted in specific experimental conditions. Thus, the maximal data transmission of the classic channel with leaked $0.17\%$ photons in the transmission channel is 0.0017 bit per detection. However, we achieved 0.8 bit per detection with $D_0$ and $D_1$ on average. Thus, the real information transmission rate is significantly higher than 0.0017 bit per detection, which demonstrates the counterfactual property during our transmission.\par

Via the quantum Zeno effect, our work implements the modified protocol with a long interference chain of outer and inner cycles via quantum nanophotonic chip, which guarantees the interference visibility and high success probabilities for bit 0 and bit 1. Benefiting from integrated circuit manufacturing, packaging technology, and booming quantum detection system, we adopt the integrated scheme to accomplish this experiment with lower error rate, compared with those implementations in free space~\cite{cao2017direct,pan2023counterfactual}. Moreover, to practically verify the counterfactual property, $D_f$ is appended at the end of the inner cycles to monitor photons' leakage. Thus, the modified weak-trace-free counterfactual communication based on the quantum Zeno effect is fully realized. Furthermore, compared with the other implementation of modified counterfactual communication protocol published recently~\cite{pan2023counterfactual}, our work implements the quantum Zeno effect, reduces the transmission time cost, and accelerates the transmission speed.

For future work, it is worth considering the practical scalability limits of the nested interferometer architecture as $M$ and $N$ increase, along with potential strategies to mitigate these challenges. First, with $M$ and $N$ increasing, leading to a larger chip footprint and a reduced success yield~\cite{qiang2021implementing}, which is addressable through compact component designs, advanced wafer-scale manufacturing, or employing distributed chip architectures. Second, thermal crosstalk between adjacent thermo-optic PSs can be reduced using optimized bias calibration algorithms~\cite{bao2025ultra,Lin:24}, cascaded sub-array designs, or electro-optic modulation techniques. Third, control complexity is manageable through intelligent control algorithms and time-division multiplexing architectures~\cite{Lin:24}. These approaches collectively offer pathways toward scaling this architecture for future large-scale implementations.

\section{Method}
\label{sec:Method}

\begin{figure*}[t]
    \centering 
    \includegraphics[width=0.95\textwidth,height=0.43\textwidth]{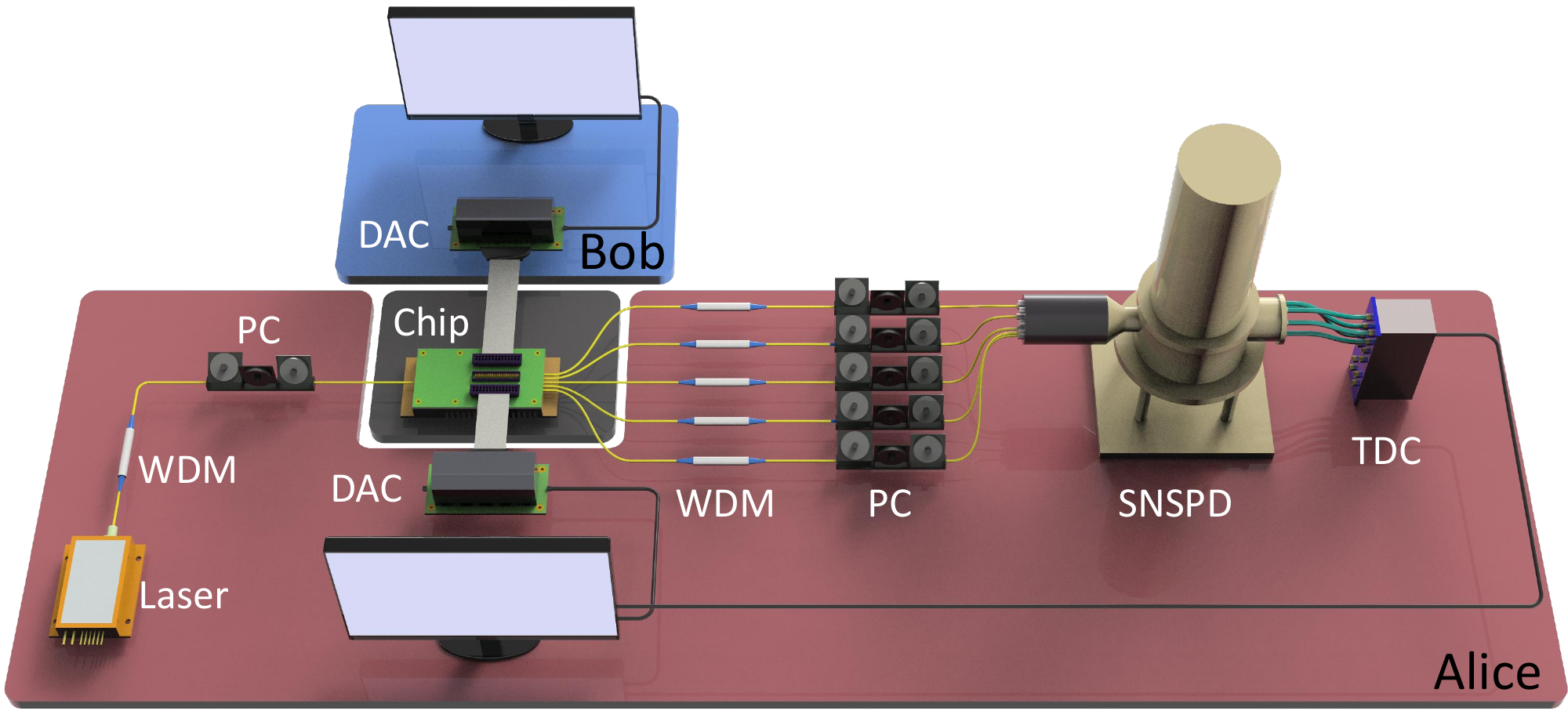}  
   \caption{\justifying{{\bfseries The whole experiment platform.} Before entering the chip, pump light is produced by the laser and is adjusted with several steps, including spectral filtering by WDM and polarization modification by PC. The parameters of the on-chip interferometer are adjusted by the signal DAC with Alice and Bob. Finally, the output photons are adjusted again and are detected by the SNSPD and counted with the TDC.}}
   \label{Platform}
\end{figure*}
\subsection{The platform of our experiment}
We implement the counterfactual communication via an integrated photonic chip, taking advantage of the stable control of the tandem interferometers. This on-chip implementation is an electrical-controlled quantum photonic system, shown as Fig.~\ref{Platform}. Firstly, a tunable continuous-wave laser at \SI{1549.32}{\nano\meter} is filtered by a wavelength division multiplexing~(WDM) module and is modulated by a polarization controller~(PC), coupled to the chip as pump light. Then the ring resonator on the chip converts pump light to single-photon pairs. Two asymmetric MZIs filter the pump light and separate idler photons, acting as a trigger for the photon pairs' coincidence detection. The signal photon enters the cycled interferometers to start the progress of counterfactual communication. The interferometers are realized by MZIs and phase shifters~(PSs) on the chip, which are configured by a digital-to-analog converter~(DAC) and are controlled by a computer. Signal photons are detected by a superconducting nanowire single-photon detector~(SNSPD), whose polarization is adjusted by in-line polarization controllers. Time-to-digital converter~(TDC) records the photon's arrival and the two-photon coincidence events are recorded by the computer. The water cooling system cooperating with a thermoelectric cooler~(TEC), keeps the device temperature constant with a proportional integrative derivative controller.

\subsection{Phase compensation and loss offset}
\label{compensation}

Due to the difference in reflectivity between MMIs and waveguides, as well as the varying number of MMIs in different paths, the transmission losses of different paths are not identical. This difference in loss would be accumulated with the increased number of M and N, which limits the scalability of the scheme if no compensation is conducted. To address this issue, we first quantify path imbalances by injecting classical light and measure the output optical power for the bit 0 and bit 1 paths, identifying transmission losses of \SI{11}{\dB} and \SI{17}{\dB}, respectively. Based on these measurements, we adjust the MZI voltage to tune the effective reflectivity of the BSs, thereby compensating for the loss discrepancy. Although inherent losses vary due to differing path lengths and component counts, the pre-adjusted MZI rebalances the photon's transmitting probabilities to ensure high-visibility interference.

PSs are based on the thermo-optical effect and introduce additional phase shifts compared to waveguides. This leads to thermal crosstalk and phase shifts, which affect the interference effect. Moreover, compared to the outer loop path, the inner loop path experiences more severe phase shifts after passing through the $BS_N$ and more MMIs. Therefore, an adjustable PS is used in the inner loop path to correct the phase shift of the outer loop path, and this calibration is performed under the condition that other modulators are set to working parameters. This optimization of phase implementation ensures the accuracy of control and the precision of the success probability of information transmission in counterfactual communication. As a result, the interference visibility within the entire structure reaches 99.67\%, defined as the ratio of the interferometer's minimum output intensity to its maximum output intensity, given by $V= \frac{I_{max}-I_{min}}{I_{max}+I_{min}}$. The long-term stability of the system supports the sustained high success probabilities (74.2\% for bit 0 and 85.1\% for bit 1) over a 5-hour continuous experiment.

\subsection{The heralded single-photon source}
Relying on spontaneous four wave mixing~(SFWM), the reconfigurable silicon micro-ring resonator acts as the quantum light source~\cite{wu2022optimization}, creating a pair of idler and signal photons. In order to guarantee the spectral purity, there are two asymmetric MZIs, Filter 1 and Filter 2, to filter the pump light and idler photons, respectively. Filter 1 achieves a pump suppression ratio over \SI{20}{\dB} with pump light at C35 (\SI{1549.32}{\nano\meter}). Filter 2 utilizes wavelength-dependent interference to separate idler photons at C31 (\SI{1552.52}{\nano\meter}) from signal photons at C39 (\SI{1546.12}{\nano\meter}), directing them to distinct output ports. With high-yield ring resonator after optimized, our system produces ~\SI{2.6e7} pairs photons per second, in which signal photon is used to transmit the information and the other idler one is being as trigger of coincidence count. After~\SI{17}{\dB} transmission channel loss, $40\%$ coupling efficiency, and $90\%$ detection efficiency of the single photon detectors, as a consequence, the effective brightness of the single-photon source is about \qty[per-mode = symbol]{1.9e5}{\per\second}.

\section*{Acknowledgements} This work was supported by National Natural Science Foundation of China (62371459, 62061136011, 62105366 and 62405368), the Quantum Science and Technology—National Science and Technology Major Project (2021ZD0300704). X.Q. acknowledges support from the National Natural Science Foundation of China (Grant No.62075243), the National Natural Science Foundation of China ``Mathematical Basic Theory of Quantum Computing'' special project (Grant No.12341103), and the National Young Talents Program.

\section*{Author contributions} Anqi Huang, Ping Xu, Yizhi Wang, and Chao Wu designed the quantum photonics chip. Tianyi Xing, Anqi Huang, and Yizhi Wang conducted the experiment with help from Yaxuan Wang, Pingyu Zhu, Jiangfang Ding, Yingwen Liu. Anqi Huang, Tianyi Xing and Junjie Wu wrote the article with help from all authors. Junjie Wu and Ping Xu supervised the study.

\section*{Data availability}
The data that support the plots within this paper and other findings of this study are available from  corresponding authors upon reasonable request.

\section*{Competing  interests}
The authors declare no conflicts of interest.

\twocolumngrid

\bibliographystyle{apsrev4-2.bst}
\bibliography{main.bib}

\end{document}